\documentclass[twocolumn,floatfix]{aastex631}
\usepackage{hyperref}

\usepackage{amsmath}
\usepackage{float,soul}
\let\cleardoublepage=\clearpage

\begin{document}

\title{The first radio view of a type Ibn supernova in SN\,2023fyq: Understanding the mass-loss history in the last decade before the explosion}

\correspondingauthor{Raphael Baer-Way}
\email{bek5cw@virginia.edu}

\newcommand{\NRAO}{\affiliation{National Radio Astronomy Observatory,
520.0Edgemont Rd, Charlottesville VA 22903, USA}}
\newcommand{\UVA}{\affiliation{Department of Astronomy, University of Virginia,
 Charlottesville VA 22904-4325, USA}}
\newcommand{\UCB}{\affiliation{Department of Astronomy, University of California, Berkeley, CA 94720-3411, USA}}
\author[0009-0004-7268-7283]{Raphael Baer-Way}\UVA \NRAO
\author[0000-0002-8070-5400]{Nayana A. J.}\UCB
\affiliation{Berkeley Center for Multi-messenger Research on Astrophysical Transients and Outreach (Multi-RAPTOR), University of California, Berkeley, CA 94720-3411, USA}
\author[0000-0002-3934-2644]{Wynn Jacobson-Galán}
\altaffiliation{NASA Hubble Fellow}
\affil{Department of Astronomy and Astrophysics, California Institute of Technology, Pasadena, CA 91125, USA}
\author[0000-0002-0786-7307]{Poonam Chandra} \NRAO 
\author[0000-0001-7132-0333]{Maryam Modjaz} \UVA

\author[0000-0003-2872-5153]{Samantha C. Wu}
\affiliation{The Observatories of the Carnegie Institution for Science, Pasadena, CA 91101, USA}
\affiliation{Center for Interdisciplinary Exploration \& Research in Astrophysics (CIERA), Physics \& Astronomy, Northwestern University, Evanston, IL 60202, USA}
\author[0000-0002-6347-3089]{Daichi Tsuna} \affiliation{TAPIR, Mailcode 350-17, California Institute of Technology, Pasadena, CA 91125, USA}
\affiliation{Research Center for the Early Universe (RESCEU), School of Science, The University of Tokyo,  Bunkyo-ku, Tokyo 113-0033, Japan}
\author[0000-0003-4768-7586]{Raffaella Margutti}\UCB
\affiliation{Department of Physics, University of California, 366 Physics North MC 7300, Berkeley, CA 94720, USA}
\affiliation{Berkeley Center for Multi-messenger Research on Astrophysical Transients and Outreach (Multi-RAPTOR), University of California, Berkeley, CA 94720-3411, USA}

\author[0000-0002-7706-5668]{Ryan~Chornock}\UCB\affiliation{Berkeley Center for Multi-messenger Research on Astrophysical Transients and Outreach (Multi-RAPTOR), University of California, Berkeley, CA 94720-3411, USA}

\author[0000-0002-7472-1279]{Craig Pellegrino} \UVA

\author[0000-0002-7937-6371]{Yize Dong}
\affiliation{Center for Astrophysics $|$ Harvard \& Smithsonian, Cambridge, MA 02138, USA}
\affiliation{The NSF AI Institute for Artificial Intelligence and Fundamental Interactions}
\author[0000-0001-7081-0082]{Maria R. Drout}
\affiliation{David A. Dunlap Department of Astronomy \& Astrophysics, University of Toronto, 50 St George St, Toronto, ON M5S 3H4, Canada}
\author[0000-0002-5740-7747
]{Charles~D.~Kilpatrick}\affiliation{Center for Interdisciplinary Exploration and Research in Astrophysics (CIERA), Northwestern University, Evanston, IL 60208, USA}
\affiliation{Berkeley Center for Multi-messenger Research on Astrophysical Transients and Outreach (Multi-RAPTOR), University of California, Berkeley, CA 94720-3411, USA}
\author[0000-0002-0763-3885]{Dan Milisavljevic}
\affiliation{Department of Physics and Astronomy, Purdue University, 525 Northwestern Avenue, West Lafayette, IN 47907-2036, USA}
\author[0000-0002-7507-8115]{Daniel Patnaude}
\affil{Smithsonian Astrophysical Observatory, 60 Garden Street, Cambridge, MA 02138 USA}
\author[0000-0001-8769-4591]{Candice Stauffer}\affiliation{Center for Interdisciplinary Exploration \& Research in Astrophysics (CIERA), Physics \& Astronomy, Northwestern University, Evanston, IL 60202, USA}



\begin{abstract}
Supernovae that interact with hydrogen-poor, helium-rich circumstellar material (CSM), known as Type Ibn supernovae (SNe Ibn), present a unique opportunity to probe mass-loss processes in massive stars. In this work, we report the first radio detection of a SN Ibn, SN 2023fyq, and characterize the mass-loss history of its stellar progenitor using the radio and X-ray observations obtained over 18 months post-explosion. We find that the radio emission from 58--185 days is best modeled by synchrotron radiation attenuated by free-free absorption from a CSM of density $\sim$ $10^{-18}$ g/$\rm{cm^{3}}$ ($\sim 10^{6} \mathrm{\rho_{ISM}}$) at a radius of $10^{16}$ cm, corresponding to a mass-loss rate of $\sim$ $4 \times 10^{-3} \ \mathrm{M_{\odot} \ yr^{-1}}$ (for a CSM velocity of 1700 km/s from optical spectroscopy) from 0.7 to 3 years before the explosion. This timescale is consistent with the time frame over which pre-explosion optical outbursts were observed. However, our late-time observations at 525 days post-explosion yield non-detections, and the 3$\sigma$ upper limits (along with an X-ray non-detection) allow us to infer a drop in progenitor mass-loss rate at 5-10 years pre-explosion with $\rm{\dot{M}}$ $< 2.5\times 10^{-3} \ \mathrm{M_{\odot} \ yr^{-1}}$. These results suggest a shell-like CSM from at most $4 \times 10^{15}$ to $2 \times 10^{16}$ cm ($\sim 10^{5} R_{\rm{\odot}}$) with a CSM density that is roughly consistent with predictions from a merger model for this object. Future radio observations of a larger sample of SNe Ibn will provide key details on the extent and density of their helium-rich CSM.
\end{abstract}

\keywords{Stellar mass loss (1613) --- Core-collapse supernovae (304)  ---  Circumstellar matter (241) }

\section{Introduction} \label{sec:intro}
Massive stars lose significant amounts of mass as they approach the ends of their lives \citep{Woosley_1993,Smith_2014,Chiosi_86,Langer_2012,Puls_2008}. Observations of certain supernovae (SNe) reveal extensive interaction between the ejecta and dense circumstellar material (CSM) formed by this mass loss (see e.g., \citealt{Chevalier_17,Margutti_2017,Chandra_2018,WJG_24,Brethauer_2022, Maeda_2010jl,BaerWay2025}). In particular, certain objects reveal interaction between the ejecta and helium-rich CSM \citep{Matheson_2000,Hosseinzadeh_2017, Pastorello_2016,Moriya_2016,Foley_2006jc,Smith_2016}, suggesting the star has almost fully lost its hydrogen and lost at least some of its helium layer but experienced this mass loss close enough to the explosion to retain a meaningful amount of nearby high-density helium-rich material. 

These objects, known as type Ibn supernovae (SNe Ibn) as they are hydrogen-deficient but their optical spectra contain narrow helium emission lines ($\sim$ 1000 km/s FWHM \citep{Hosseinzadeh_2017}) from unshocked CSM \citep{GalYam2017,Modjaz_2019}, make up 1$\%$ (by local volumetric rate) of the Core Collapse Supernova (CCSN) population \citep{Perley_2020}. It has been somewhat difficult to observationally constrain this class of objects due to their relative scarcity. SNe Ibn are characterized by fast-declining (relative to any other supernova subtype) lightcurves \citep{Moriya_2016,Hosseinzadeh_2017,Khakpash}, with the $\sim$ 0.1 mag/day decay in the first month post-explosion driven by changes in the shock emission mechanism from radiatively cooling to adiabatic. This occurs in SNe Ibn due to the high shock velocities (which lower the CSM densities by quickly propagating to large radii and allowing for a faster cooling transition) characteristic of the subtype \citep{Maeda_2022}. By definition, all SNe Ibn interact with some type of helium-rich CSM, but optical spectral diversity in line-width and shape (i.e., P Cygni profiles vs. solely narrow emission lines) may suggest different optical depths or viewing-angle effects \citep{Shivvers_2017,Hosseinzadeh_2017}. Viewing-angle effects would imply that there are large deviations from spherical symmetry in the explosion/CSM itself.
\par Some SNe Ibn show narrow-line emission (order 1000 km/s speed) that persists out to late times and is indicative of continued interaction into the nebular phase $\sim$ 200--300 days post-explosion \citep{Foley_2006jc,smith08}. Others, such as the target we analyze in this paper, evolve to look more like a classical stripped-envelope SN at later phases, potentially due to a less extended helium shell of CSM \citep{Dong_2024}. Given the diversity within the subclass, it is possible that there are intrinsic differences between these explosions/their progenitors. This implies that they should not be grouped as one subclass, or that the mass-loss mechanism can act on varying timescales. The time frame over which the intensive mass-loss occurs ($>10^{-3} M_{\odot}\, \rm yr^{-1}$ as has been seen in all SNe Ibn for which a mass-loss rate was measured \citep{Pellegrino_2024}) is much shorter (on the order of years) than in hydrogen-rich interacting objects, where the mass loss can occur for hundreds of years \citep{Smith_2005ip}. There are various possibilities for these short-lived mass loss episodes of SNe Ibn, such as mass ejections due to pulsational instabilities near core collapse \citep{Woosley_2017}, or mass loss due to unstable mass transfer during inspiral in a binary system on the way to a merger-driven explosion \citep{Wu_2022,Tsuna_2024,Schroder_2020}.

\par Roughly 70--100 SNe Ibn have been discovered at optical wavelengths \citep{Dong_2024}. Of these, three have shown evidence for pre-explosion outbursts, detected as flares or excess emission in pre-explosion lightcurves \citep{Gangopadhyay_2020,Foley_2006jc,Brennan_2024}, one of these being the object we explore in this work. Beyond SNe Ibn, there have been a few tens of cases of precursor emission observed in CCSNe \citep{Strotjohann_2021,Pastorello_2013,ER_2016,Ofek_2014}. In most cases, including the Ibn subclass, these precursors are interpreted as compelling evidence for mass ejection before the explosion that creates dense CSM.
\citep{Gagliano_2025,WJG_22,Bilinski_2015,Ofek_2013,Pastorello_2010}. 

As the supernova explodes, the ejecta will run into the CSM, causing non-thermal radio synchrotron emission given the strong magnetic field and fast ($\sim$ 10,000 km/s) shock \citep{Chevalier_1982}. The hot shocked material can also generate thermal X-rays \citep{Chevalier_1982}. Three type Ibn SNe have been detected at X-ray wavelengths: SN 2006jc \citep{Immler_2008}, SN 2010al \citep{Ofek_2013} (initially classified as a type IIn) and SN 2022ablq \citep{ Pellegrino_2022}. These X-ray observations revealed that these SNe likely had an intense outburst of mass-loss $\sim$ 1-2 years pre-explosion. While roughly 100 CCSNe have been detected at radio wavelengths \citep{Bietenholz_2021}, a SN Ibn has never been detected despite attempts for multiple different objects \citep{Shivvers_2017,smith08}. Given that very few objects have been sampled out to late times (beyond $\sim$ 30 days-meaning the mass loss was only probed out to $\sim$ 1 year pre-explosion given characteristic 10,000/1000 km/s shock/wind speeds) and/or with high sensitivity $< 10^{26}\hspace{0.1 cm}\rm{ergs\hspace{0.1 cm} s^{-1}\hspace{0.1 cm} Hz^{-1}}$, these non-detections are not particularly constraining for the subclass as a whole.
\par In this work, we report the first radio detection of an SN Ibn, SN 2023fyq. We use the radio emission of SN\ 2023fyq to constrain the mass-loss history of its progenitor system. SN\ 2023fyq has been studied extensively at optical wavelengths even before explosion: SN\, 2023fyq showed evidence for optical precursor emission indicative of pre-explosion outbursts for at least 1500 days \citep{Dong_2024,Brennan_2024}. The object then rose dramatically in brightness, and the explosion date (in our frame) was inferred as 16 July 2023 \citep{Brennan_2024}. \citet{Brennan_2024} estimated this explosion date relative to the time of maximum through blackbody fits. We take all epochs in this work relative to this explosion date. We note that our results are robust even with an explosion date error of $\sim$ $\pm$ 15 days. The SN is roughly 15" offset from the center of its host galaxy, NGC\,4388 -- which is at a distance of 18 $\pm$ 3.7 Mpc using the most recent Tully-Fisher estimate \citep{Dong_2024}. This galaxy has attracted interest on its own as a unique radio and X-ray-bright active galactic nucleus (AGN) \citep{Sargent_2024,Goldman_2024}.

\par  SN\,2023fyq is one of the first objects for which pre-explosion spectra were acquired of the outbursting material \citep{Brennan_2024}. \cite{Dong_2024} and \cite{Brennan_2024} have combined the optical pre- and post-explosion spectra and lightcurves (which showed that this SN was unambiguously a terminal transient-meaning the core was disrupted and the progenitor destroyed) to characterize SN\,2023fyq. They interpreted their optical dataset of this object as an explosion originating from a binary system with a low-mass ($\sim$ 3 $M_{\odot}$) helium star orbiting a neutron star that they claim eventually experienced either runaway mass loss triggering core collapse, or a merger that directly triggered the explosion \citep{Dong_2024}.

\par 
We present the radio and X-ray observations of SN\,2023fyq to study the CSM of the progenitor system. The layout of the paper is as follows: in \S \ref{sec:DR}, we describe the radio and X-ray observations and data reduction processes.
In \S \ref{sec:DA} we present the modeling of the data in a CSM interaction scenario. In \S \ref{sec:Discussion}, we present the results, discuss their implications, and compare our findings with the results obtained at optical wavelengths. Conclusions are drawn in \S \ref{sec:Conclusions}.
\section{Observations and Data Reduction}\label{sec:DR} 
\begin{figure}
    \centering
    \includegraphics[width=8 cm, height=6 cm]{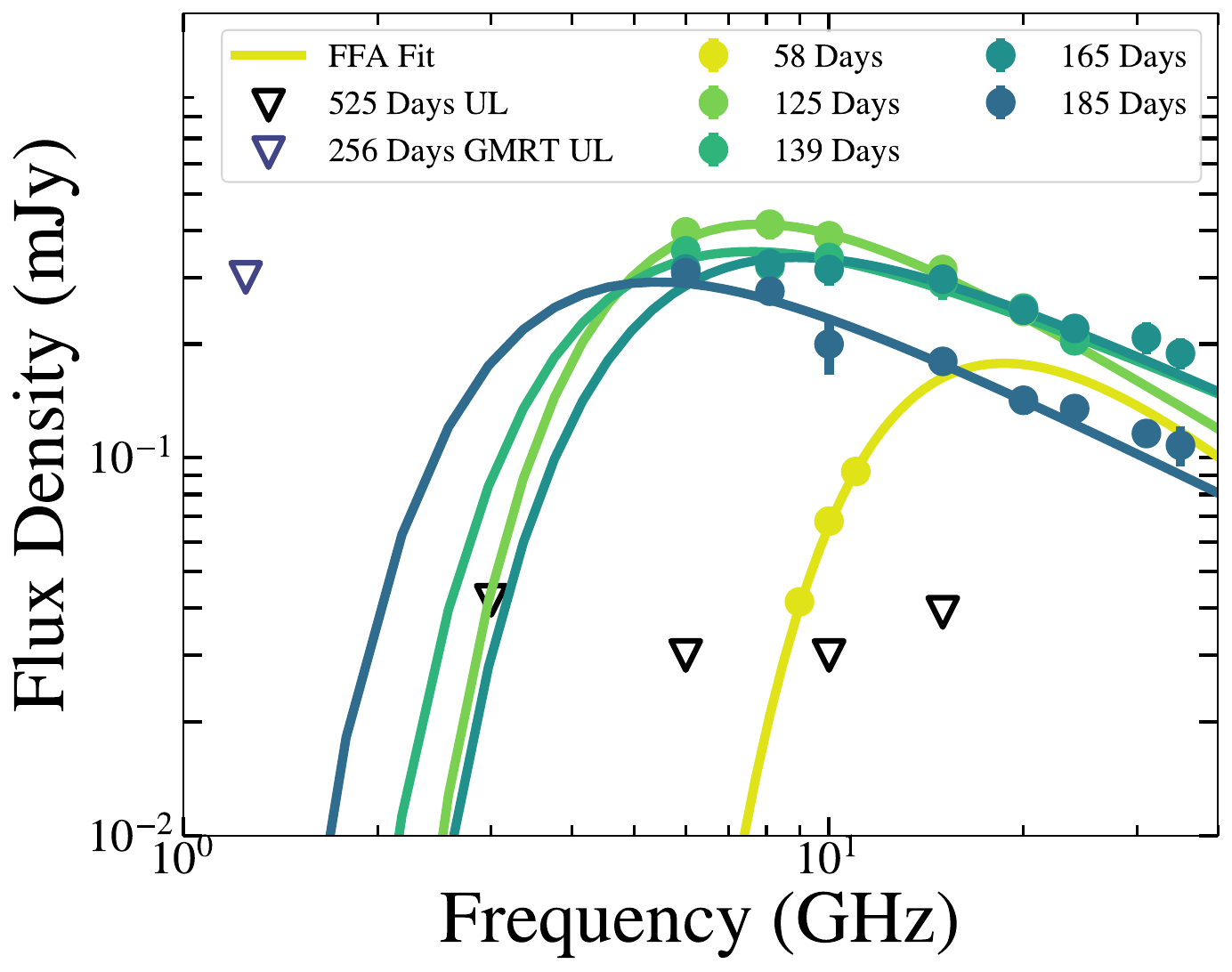}
    \caption{Radio SEDs of SN\,2023fyq at VLA frequencies (3--35 GHz) spanning 58--525 days post explosion. We show the best-fit extrapolated single-epoch FFA models for each epoch as described in section \ref{sec:DA}. Downward-facing triangles denote 3$\sigma$ flux density upper limits. The parameters for these fits are in Table \ref{tab:2023fyq_Fitting_results}.}
    \label{fig:Raw_Dat}
\end{figure}
\subsection{GMRT}
We obtained Giant Meterwave Radio Telescope (GMRT) observations of SN\,2023fyq. We carried out GMRT observations on 12 August 2023 ($\Delta$t= 27 days) and 28 March 2024 ($\Delta$t= 256 days). Both were obtained with program 45-091 (PI P. Chandra). Both observations were in band 5 (1060--1460 MHz) with a total integration time of two hours. We used a bandwidth of 400 MHz split into 2048 channels each with a width of 195.3 kHz. We used 3C147 as the flux and bandpass calibrator, and J1120+143 as the phase calibrator. We reduced the GMRT data using the CAPTURE pipeline \citep{GMRT_pipeline}, though we did not detect the SN following repeated self-calibrations within the pipeline. At band 5, the GMRT has angular resolution $\sim$ 2 arcsec$^{2}$. Given the relatively high resolution, the nearby AGN host (at 15\hbox{$^{\prime\prime}$} from the SN) did not contaminate the emission from the SN location despite the bright AGN radio emission ($\sim 5$ mJy)  at 1.25 GHz. We obtained 3$\sigma$ upper limits using the RMS in a region multiple times the beam size at the SN location at both epochs as reported in Table \ref{tab:2023fyq}.
\par 
\subsection{VLA}
 We obtained 8 epochs of Karl G. Jansky Very Large Array (VLA) data spanning from July 22, 2023 to December 23 2024 ($\Delta$t=6 to 525 days) of SN\,2023fyq. The VLA data up to 185 days were obtained with program VLA/23A-157 (PI W. Jacobson-Gal\'{a}n) and the late-time epoch was with VLA/24B-500 (PI
R. Baer-Way). The flux calibrator 3C286 and phase calibrator J1223+1141 were used for each observation. Five of the epochs were multi-frequency observations covering S (2--4\, GHz), C (4--8\,GHz), X (8--12\,GHz), Ku (12--18\, GHz), K (18--26.5\,GHz), and Ka (26.5--40\,GHz) bands, and 3 of these were X band snapshots. The initial X band snapshots and the final epoch at $\sim$ 1.5 years post-explosion were taken in the VLA A-configuration. We identified the SN in all radio images by cross-referencing with the known optical location of the source \citep{Brennan_2024}.  The four epochs of multi-frequency observations were taken in quick succession following the initial detection in September of 2023 ($\Delta t=58$ days) at 10 GHz at $\sim$ 0.1 mJy.  These multi-frequency epochs were taken in D or D $\rightarrow$ C configuration with low angular resolution (i.e. at S band the beam is $\sim$ 4.6 $\times$ 4.6 arcsec$^{2}$). 

Proximity to the radio-bright host galaxy (with size 6.2 $\times$ 1.4 arcmin$^{2}$ at $\sim$ 4 mJy in C band \citep{Sargent_2024}) made data reduction difficult for D/C configuration data. 
We reduced the VLA data using standard Common Astronomy Software Applications (CASA) \citep{Casa_desc} techniques with custom flagging. 
We performed multiple rounds of self-calibration on the low-frequency data (S/C/X band). To further address the contamination from the host galaxy, we made stringent cuts using CASA's \texttt{tfcrop} to remove noise at a 2$\sigma$ level, and then removed all data taken with UV baselines $< 5000\lambda$ at each associated $\mathrm{\lambda}$. This ensures that we remove data corresponding to extended emission from sources of $>0.6$ arcminutes in size and thereby remove extended emission from the host. This isolated the point sources near the host in the image and allowed for robust flux fitting using CASA's \texttt{imfit} at 6 GHz and higher frequencies \citep{Casa_desc}. 
 To confirm the emission, we also used images of the host taken prior to the explosion in C band to carefully calibrate and subtract, allowing a second check on the bright emission. We find agreement with our results. We report the flux density values at the SN position found using \texttt{imfit} in table \ref{tab:2023fyq}.
For model fitting, we add $10\%$ of the flux in quadrature to the errors from the image fitting to account for
systematic uncertainties \citep{Weiler_1986} and the fact that even after careful reduction, the flux density of the SN is slightly variable (at the $\sim$ 5\% level) with different clean iterations/weighting. At S band (3 GHz), the SN could not be resolved from the host galaxy and thus we only report $3\sigma$ flux density upper limits."

\subsection{Swift-XRT Observations} 
SN\,2023fyq was observed with the X-ray telescope (XRT) onboard the Neil Gehrels \emph{Swift} Observatory \citep{Gehrels04} across 30 epochs post-explosion, all shorter than 2\,ks. We binned these observations into 2 subgroups with larger exposure times, with one at $\Delta t=3 \pm 1$\,d (10.5 ks observation of the SN) and the other at 160 $\pm 7$ days (combined 17.7 ks observation, where we grouped observations from ObsID 37262-- observations of a nearby maser $\sim$ 10\hbox{$^{\prime\prime}$} away and the SN itself). Additionally, the host galaxy was observed 1 month before the explosion for 4 ks (ObsID 32364).  We reduced all data using the HEAsoft \texttt{xrtpipeline} \citep{HEAsoft}, which performs all relevant screening and processing steps. However, NGC 4388 contains the aforementioned X-ray bright AGN \citep{Goldman_2024} that contaminates the X-ray flux from the SN location in all epochs of Swift-XRT observations. As shown by \cite{Goldman_2024}, the continuum flux of NGC 4388 changes by a factor $\sim$ 2 over a period of 4 years. Given the slow variability of the AGN, we use the observation obtained two months prior to the SN  to estimate the ''background" AGN emission. We opt to use the epoch directly 2 months pre-explosion to attempt to constrain the underlying AGN flux at the SN location.
\par
\begin{figure}
    \centering
    \includegraphics[width=8 cm, height=6 cm]{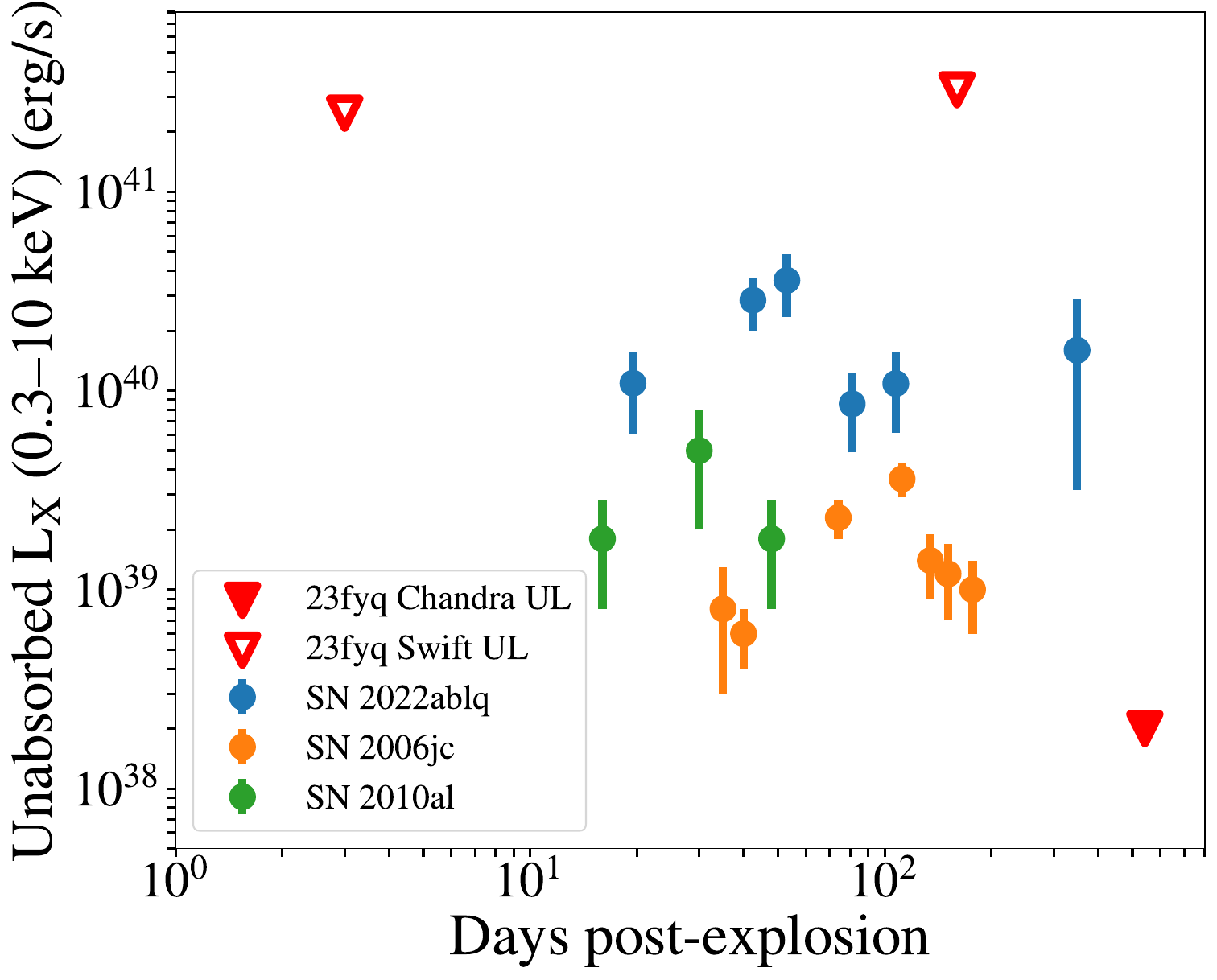}
    \caption{The unabsorbed X-ray light curves of detected SNe Ibn, along with the non-detections of SN 2023fyq. \textit{Swift} Data on SN 2006jc, SN 2010al and SN 2022ablq are from \citet{Immler_2008,Ofek_2013,Pellegrino_2024}.  Downward triangles represent 3$\sigma$ upper limits from \textit{Chandra} and \textit{Swift} for SN 2023fyq. Deeper, earlier observations would be vital to constrain the subclass better at X-ray wavelengths.}
    \label{fig:Xray_fig}
\end{figure}

  We used a 10\hbox{$^{\prime\prime}$} region centered around the SN (to try to avoid the nucleus of the AGN) and extracted a spectrum (estimating the background in a large region of radius $\sim 100 \hbox{$^{\prime\prime}$}$ that is source-free) at both combined epochs and the pre-explosion epoch. We find a relatively constant 0.3--10 keV count rate in the region (within 10$\%$) at both pre- and post-explosion epochs. We fit an absorbed power law (a phenomenological AGN model) to the pre-explosion data, finding $\Gamma=1.47 \pm 0.22$ and column density $(3.2 \pm 1.05) \times 10^{23} cm^{-2}$. We then fit the post-SN data with a model consisting of two components: an AGN absorbed power-law from the pre-SN fit and another absorbed power-law model that represents the SN flux. Given we are using nested models, we perform an F-test by calculating the F-statistic given the change in $\chi^2$ from adding the second model and the added degrees of freedom. We find an F-statistic 
$\sim$ 1.5 at both epochs, which, given the degrees of freedom and added parameters, implies a p-value $> 0.05$. We thus find no statistical evidence at either post-explosion epoch for flux from the SN. We find a 3$\sigma$ upper limit on the 0.3--10 keV flux of this SN component by adding a \texttt{cflux} component as part of the SN model and freezing all other parameters (using \texttt{steppar} to find the 3$\sigma$ limit). We find an upper limit of $6.43 \times 10^{-12} \rm{erg \hspace{0.1 cm} cm^{-2}\hspace{0.1 cm} s^{-1}}$ at the first epoch, and $8.43\times10^{-12} \rm{erg \hspace{0.1 cm} cm^{-2}\hspace{0.1 cm} s^{-1}}$ at the second epoch. We show these upper limits in Figure \ref{fig:Xray_fig}. We conclude that the Swift-XRT data show no conclusive evidence for significant X-ray emission from SN\,2023fyq.
\subsection{\textit{Chandra} Observations}
\par The \textit{Chandra X-ray Observatory (CXO)} observed SN\,2023fyq on 8 January 2025 (PI Y. Dong, proposal number 5509011, $\Delta t=542$ days). With the high angular resolution of \textit{CXO} ($\sim$0.5\hbox{$^{\prime\prime}$}), the SN position is not contaminated by the host galaxy. We did not detect X-ray emission from the SN. We extracted the data using standard {\tt CIAO} techniques, and redid astrometry to ensure we can compare directly to the optical position of the SN. We used the \textit{Gaia} \citep{Gaia_2023} catalog to do astrometry, using 3 sources (due to the low number of sources detected in the \textit{Chandra} image) and obtaining an astrometric uncertainty of 0.92\hbox{$^{\prime\prime}$}. Using the \texttt{aplimits} tool in {\tt CIAO} \citep{Ciao_2006}, we find an upper limit on the count rate at the SN position of $< 2.55\times 10^{-4} \rm{counts/s}$ (in the 0.5--8 keV band ). We use a 3 arcsecond region for the SN and a 50 arcsecond source-free region for the background to obtain this upper limit on the count rate. To attempt to quantify whether we are capturing any flux from the host galaxy that could contaminate the SN, we ran \textit{srcflux} to obtain the 0.5-7 keV count rate from the host galaxy from a 10 arcsecond region not containing the SN and a 15 arcsecond region that does. We find that the count rates are consistent from these two regions within error bars, suggesting that there is no emission from the galaxy at the SN location (which is also clear from visual inspection of the data).

In order to convert our upper limits on X-ray-counts into an upper limit on X-ray luminosity, we need to make a few assumptions. Here we outline our assumptions and quantify the uncertainty of our assumptions onto the derived Xray upper limits. We assume SN~2023fyq is similar to other interacting SNe with bremmstrahlung emission at T=5 keV \citep{Brethauer_2022} and column density given by the galactic line of sight absorption at 2$\times 10^{20} \ \rm{cm^{-2}}$ \citep{Guver_2009}. We note that assuming $\rm{T}=10$ keV changes the derived flux by $< 5\%$ and even assuming a column density three orders of magnitude higher (at $10^{23} \ \rm{cm^{-2}}$-much higher than expected even for strongly interacting SNe)  changes the flux by a factor $<3$. We find a limit on the X-ray 0.3--10 keV luminosity $L_{X}<2 \times 10^{38}$~ergs/s. This is at least an order of magnitude lower than all detections for SN 2006jc/SN 2022ablq as seen in Figure \ref{fig:Xray_fig} and lower than detections for most X-ray bright SNe \citep{Dwarkadas_2025}. However, this non-detection is at later times than any detection for a SN Ibn. We interpret this limit in Section \ref{sec:DA}.

\section{Radio Modeling}\label{sec:DA}
\begin{figure}
    \centering
    \includegraphics[width=8 cm, height= 6cm]{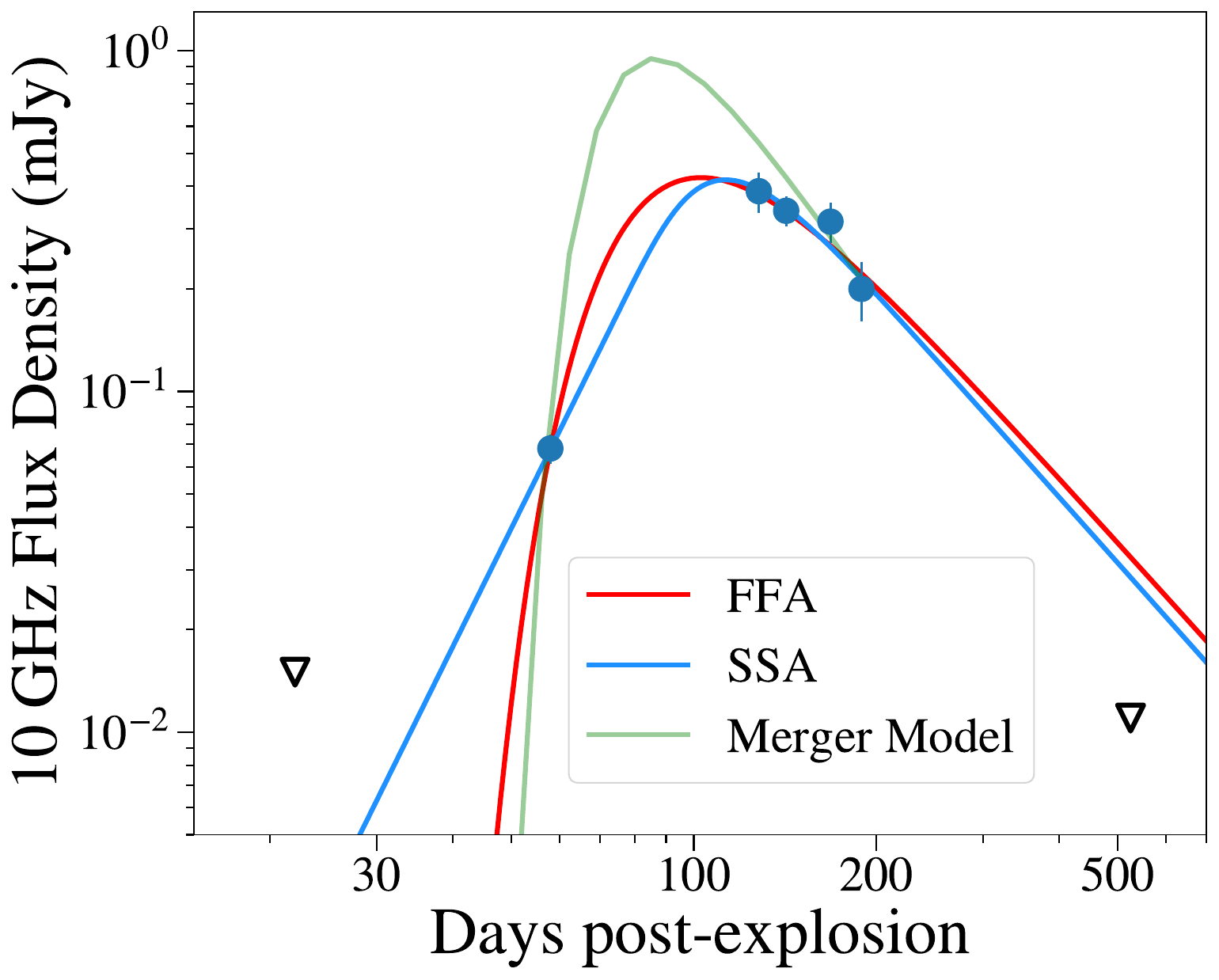}
    \caption{Radio 10 GHz lightcurve of SN\,2023fyq with various model fits. We show the best-fit Synchrotron Self-Absorbed (SSA) and Free-Free Absorbed (FFA) models at 10 GHz (where these are fits from equations \ref{eq:FFA} and \ref{eq:SSA} to the entire dataset derived in section \ref{sec:DA}) along with a model taken from \citealt{Wu_2025} (``Merger Model'') described in section \ref{sec:intrinsic SSA+FFA}. This merger model has a similar shock speed $\sim$ 9000 km/s but a different $\alpha=0.55$, and slightly different $\epsilon_{e}$  and $\epsilon_{B}$. The merger model cuts off at $\sim$ 220 days as this is when \citet{Tsuna_2024} constrain the mass loss to begin for these merger models. The 3$\sigma$ upper limits in black are not taken into account for the fits for any of the models. }
    \label{fig:10GHz}
\end{figure}
\begin{figure*}
    \centering
    \includegraphics[width=17 cm, height= 9 cm]{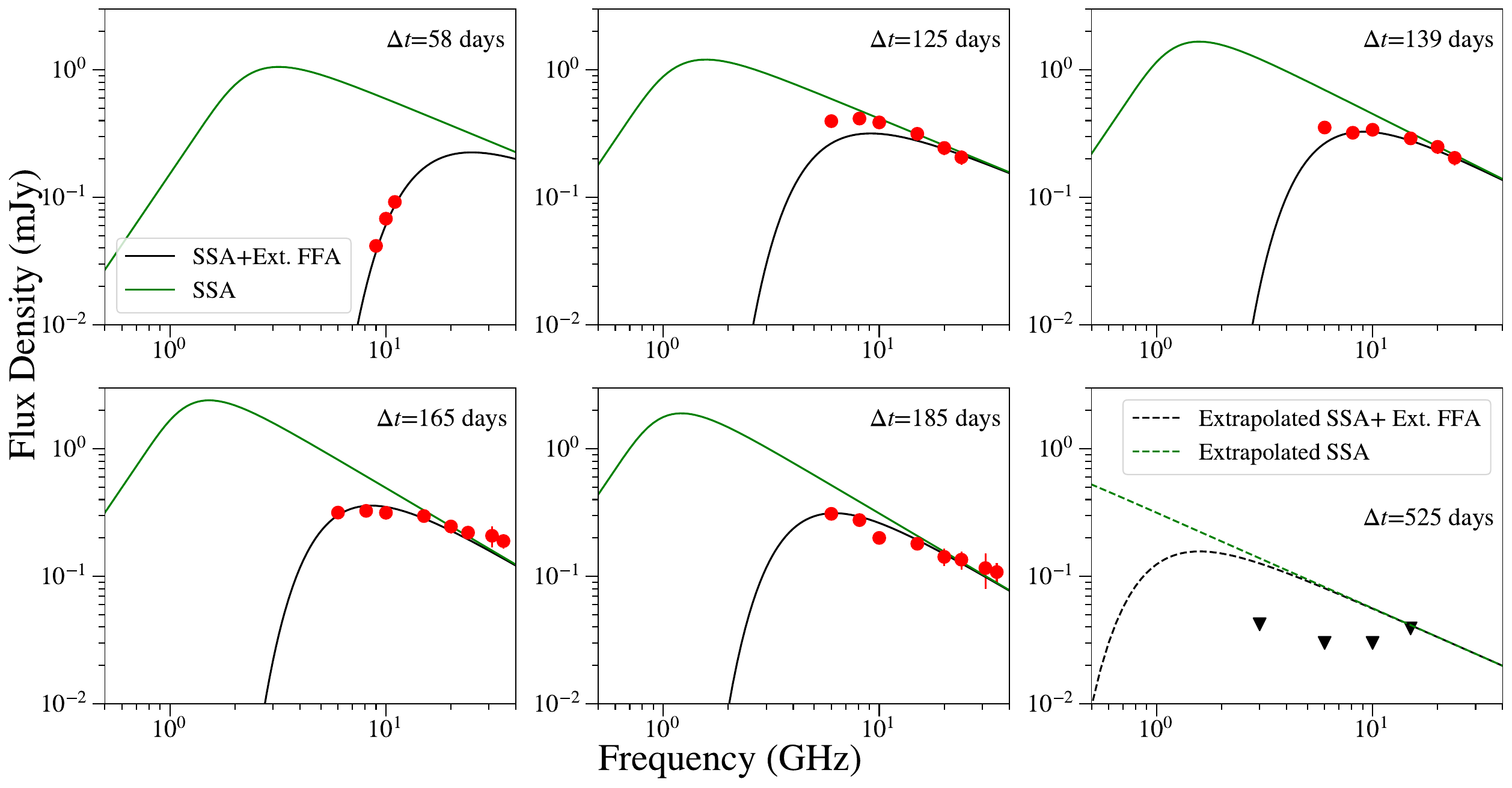}
    \caption{Radio SEDs of SN\,2023fyq at $\Delta t \approx 58-525$\,days. Downward triangles represent 3$\sigma$ flux density upper limits. Solid black lines denote the overall SSA+ external FFA model we derive self-consistently (see section \ref{sec:intrinsic SSA+FFA}). Solid green lines denote the intrinsic SSA model using the calculated $\rm{F_{p}/\nu_{p}}$ from equation \ref{eq:Sinep_SSA}.  The values of the best-fit parameters are displayed in Table \ref{tab:2023fyq_Fitting_results}. We also show the extrapolated models with dashed black and green lines at $\Delta t \approx 525$\,days with the flux density upper limits to emphasize the discrepancy between the models and the non-detections, indicating a decrease in the mass-loss rate at $\sim$ 5-10 years before the explosion. }
    \label{fig:SSAFFA_Fits}
\end{figure*}
In general, radio emission from CCSNe is non-thermal synchrotron radiation that arises due to the interaction of the SN forward shock with the CSM. The emission is suppressed via synchrotron self-absorption (SSA) by the synchrotron-emitting electrons themselves and/or via free-free absorption (FFA) by the dense medium external to the shock \citep{Chevalier_1998}.
Both absorption mechanisms can work in tandem, and radio light curves can be represented in parametrized forms that describe the optically thick ($F_{\nu}$-rising) and thin ($F_{\nu}$-declining) spectral and temporal evolution, as well as the optical depth normalization.
In an FFA model, the radio flux density ($F(\nu,t)$) can be parametrized as \cite{Chevalier_1982,Chandra_2020,Weiler_2002}:
\begin{equation}
F(\nu,t)_{\rm{FFA}} = K_{1} \left( \frac{\nu}{5 \hspace{0.1cm} \mathrm{GHz}} \right)^{-\alpha} 
\left( \frac{t}{\mathrm{1000 \hspace{0.1cm} days}} \right)^{-\beta} 
e^{-\tau_{\mathrm{FFA}}(\nu,t)} ,
\label{eq:FFA}
\end{equation}
Here, $\alpha$ is the spectral index and $\beta$ is the temporal index of the flux density evolution. $\alpha$ is related to the electron energy index $p$ by $p=2\alpha+1$ \citep{Chevalier_1998}. We note that this is true only when the self-absorption frequency is less than both the cooling and minimum electron energy frequency, but we confirm this is the case using the scale of the B-field ($\sim$ 0.5 G) derived self-consistently in section \ref{sec:intrinsic SSA+FFA}.
The FFA optical depth is given by \begin{equation}\tau_{\rm{FFA}}(\nu,t)=K_{2}\left(\frac{\nu}{5\hspace{0.1 cm} \mathrm{GHz}}\right)^{-2.1}\left(\frac{t}{1000 \hspace{0.1 cm}{\rm days}}\right)^{-\delta}\label{eq:FFA_tau}\end{equation} 
$\delta$ describes the evolution of the optical depth, and is related to the density gradient in  the CSM $\rho_{\rm{CSM}}\propto r^{-s}$ by $m=\frac{n-3}{n-s}$ (where the ejecta density profile $\rho_{\rm{ejecta}}\propto r^{-n}$), where $\delta = m(2s-1)$. We can also use this $m$ to characterize the evolution of the CSM radius $R \propto t^{\rm m}$. $K_{2}$ is the optical depth normalization. 

In the SSA model, the flux density is parametrized as \citep{Weiler_1986,Chevalier_1998} \begin{align}
F(\nu,t)_{\rm{SSA}} =\;& K_{1}\left(\frac{\nu}{5 \hspace{0.1 cm} \mathrm{GHz}}\right)^{2.5}
\left(\frac{t}{\rm{1000 \hspace{0.1 cm} days}}\right)^{-\beta'} \nonumber \\
&\times \left(1 - e^{-\rm{\tau_{SSA}(\nu,t)}}\right)
\label{eq:SSA}
\end{align}

The SSA optical depth is parametrized by \begin{equation}\tau_{\rm{SSA}}=K_{2}\left(\frac{\nu}{\mathrm{5 \hspace{0.1 cm} GHz}}\right)^{(-\alpha-2.5)}\left(\frac{t}{\rm{1000 \hspace{0.1 cm} days}}\right)^{(-\beta'+\beta)}\label{eq:SSA_tau}\end{equation} 
$K_{2}$ corresponds to the SSA optical depth normalization.
$\beta'$ and $\beta$ denote the temporal evolution of flux densities in the optically thick ($F \propto t^{\beta^{'}}$) and thin phase ($F \propto t^{\beta}$), respectively. 
The SSA model can be rewritten at a specific temporal epoch to reduce the model to 3 free parameters and measure their evolution as: \citep{Chevalier_1998,Soderberg_03bg}
\begin{equation}
F_{\nu, \rm SSA} = 1.582 F_{\rm p} \left( \frac{\nu}{\nu_{\rm p}} \right)^{5/2} \left( 1 - \exp \left[ -\left( \frac{\nu}{\nu_{\rm p}} \right)^{-2.5 - \alpha} \right] \right)
\label{eq:Sinep_SSA}
\end{equation}
 where $\alpha$ again is the optically thin spectral index, $\nu_{\rm{p}}$ is the frequency at which the SSA spectrum peaks (frequency at which $\rm{\tau_{SSA}} \approx 1$), and $F_{p}$ is the corresponding peak flux density.
$\nu_{\rm p}$ and $F_{\rm p}$ can be used to estimate the shock radius ($R$) and magnetic field ($B$) \citep{Chevalier_1998}. 
For $p\sim 2.5$ (given the optically thin spectral slope that we measure as detailed later in this section), we find \citep{Pacholczyk} 
\begin{equation}
    \begin{split}
    R=3.7 \times 10^{15}f_{eB}^{-1/18}\left( \frac{f}{0.5}\right) ^{-1/18}\left(\frac{F_{p}}{\mathrm{Jy}}\right)^{17/36}\times
    \\ \left(\frac{D}{\mathrm{Mpc}}\right)^{17/18}\left (\frac{\nu}{5 \hspace{0.1 cm}\mathrm{GHz}}\right)^{-1} \mathrm{cm}
    \end{split}
    \label{R_eq}
\end{equation}
\begin{equation}
    \begin{split}
    B=0.70 f_{eB}^{-2/9}\left(\frac{f}{0.5}\right)^{-2/9}\left (\frac{F_{p}}{\mathrm{Jy}}\right)^{-1/9}\times\\ \left(\frac{D}{\mathrm{Mpc}}\right)^{-2/9}\left(\frac{\nu}{5 \hspace{0.1 cm} \mathrm{GHz}}\right) \mathrm{G}
    \end{split}
    \label{B_eq}
\end{equation}
where $f_{eB}=\frac{\epsilon_{e}}{\epsilon_{B}}$ denotes the ratio between the energy density of relativistic electrons ($\epsilon_{e}$) and the magnetic field energy density ($\epsilon_{B}$). f is the volume filling factor, which we assume to be 0.5 as is standard for radio SNe with n $\sim$ 10 and s $\sim$ 2 \citep{Weiler_1986} -- although other values could be considered up to f=0.1 or 1 (considering the asymmetry seen in the optical dataset \citep{Dong_2024}) without changing  R and B by more than 10 $\%$. $D$ is the angular distance to the SN.
\par  As seen in Figure \ref{fig:Raw_Dat}, the data do not cover the peak of the SED at any epoch: at 58 days, the entire SED is in the optically thick phase, and the data are optically thin at all other epochs. 
We still attemped to fit an FFA-dominated and SSA-dominated model to the full dataset using equations \ref{eq:FFA} and \ref{eq:SSA}. We carry out fits using \texttt{emcee} \citep{Foreman_2013}, with 200 walkers and 10000 iterations. In the FFA model, we kept $\alpha,\beta,\delta,K{1},K_{2}$ as free parameters, while for SSA we kept $\alpha,\beta,\beta',K_{1},K_{2}$ as free parameters.
We find similar reduced $\chi^2$ $\rm{\chi_{\nu,FFA}}^2=1.41$ and $\rm{\chi_{\nu,SSA}}^2=1.81$. We find $\alpha=0.87_{-0.10}^{+0.10}$ for FFA and $\alpha=0.76_{-0.03}^{+0.02}$ for SSA. We also find for FFA that $\delta=2.11_{-0.08}^{+0.15}$. This $\delta$ value implies that $s \sim$ 2 for reasonable $m$ values from 0.6--1 with the relation $\delta=m(2s-1)$. We find $\beta'=3.67_{-0.30}^{+0.23}$ for SSA. Finally, we have $\beta=1.32_{-0.2}^{+0.18}$ for SSA and $\beta=1.86_{-0.14}^{+0.09}$ for FFA. From the similar $\chi^{2}$ values and visual inspection of the best-fitting models, we conclude that we cannot favor a pure FFA or SSA model for the full dataset but find preference for $s \sim$ 2.

However, the fact that at $\Delta t = 58$\,days,  the SED is best sampled in the optically thick regime allows us to constrain the dominant absorption process.  
 The flux densities rise by a factor $\sim$ 2 between 9--11 GHz, suggesting a very steep optically thick slope of $F_{\nu} \sim \nu^{4}$. In an SSA model, the optically thick spectral slope would be $F_{\nu} \propto \nu^{2.5}$ (given the ordering of cooling/electron/SSA frequencies we confirmed) \citep{Rybicki_lightman_79}. The observed optically thick spectral slope is steeper than this and suggests that FFA must be dominant at this epoch. We thus fit each SED with equation \ref{eq:FFA} (fixing t at each epoch). 
 
 We freeze $\delta=3$ (as $\delta=(2s-1)m$) as we assume constant shock velocity $m=1$ and $s$=2. We note that with $s=2,m=\frac{n-3}{n-s}$ can only be approximately 1, but we assume $m=1$ for simplicity and given the poor constraint on s. We also fix $\beta=1.86$ from the FFA fits to the full dataset as $\beta$ should not vary across individual epochs (as it sets the temporal evolution). As one further motivation for fitting with FFA, we note that given the 10 GHz peak, a Chevalier diagram \citep{Chevalier_1998} as shown in Figure \ref{fig:23fyq_Lpnup} suggests a shock speed $\sim$ 4000 km/s, which is a clear underestimate based on optical estimates, meaning SSA alone is not the primary absorption mechanism and there is a significant contribution from FFA.  We use $V_{\rm sh} = 8500\, \rm km\,s^{-1}$ from the late-time optical spectra \citep{Dong_2024}.
This likely corresponds to the fastest ejecta speed, which is not the same as the shock speed and is often an underestimation \citep{Chevalier_17}. We proceed assuming a non-decelerating shock speed, given that we lack the data to accurately model the evolution of this speed.

We perform FFA fits at each epoch using equation \ref{eq:FFA} keeping $K_{1}$, $K_{2}$ and $\alpha$ as free parameters. The results for these three parameters are shown in Table {\ref{tab:2023fyq_Fitting_results}. We note the relatively high $\alpha$ at early epochs, but also emphasize that the error bars are quite large and there are degeneracies in the fitting. We show the single-epoch FFA fits to the data in Figure \ref{fig:Raw_Dat}.
We caution that our results in general are only approximations given the $s$ and $m$ we assume.
\subsection{Intrinsic SSA+FFA}
\label{sec:intrinsic SSA+FFA}
We proceeded to find the intrinsic SSA emission suppressed by the FFA self-consistently by estimating the radius and B-field accounting for the density profile. We use the CSM densities from single epoch FFA fits.
We estimate $\tau_{\rm FFA}$ using equation {\ref{eq:FFA_tau} with the best-fit $\mathrm{K_{2}}$ and assumed $\delta$=3. The CSM density is then given by 
$\rho_{\rm FFA}=\frac{\dot{M_{\rm FFA}}}{4\pi r^2v_{\rm w}}$, where \citep{Chandra_2020}
\begin{equation}
\begin{split}
\frac{\dot{M}_{-3,\mathrm{FFA}}}{\rm{v_{CSM,1}}} = 7.5 \times 10^{-2} 
\left( 
\frac{ \mathrm{\tau_{\rm eff/FFA}} }{ 1 + \left[ \frac{ \mathrm{n(He)} }{ \mathrm{n(H)} } \right] } 
\right)^{0.5}
\left( \frac{ \mathrm{\nu} }{ \mathrm{GHz} } \right)^{1.06} \\
\times \left( \frac{ \mathrm{T_{\rm CSM}} }{ 10^{4} \, \mathrm{K} } \right)^{0.67}
\left( \frac{ \mathrm{v_{sh}} }{ 10^4 \, \mathrm{km\,s^{-1}} } \right)^{1.5}
\left( \frac{ \mathrm{t} }{ 1000 \, \mathrm{days} } \right)^{1.5}
\end{split}
\label{eq:Mdot}
\end{equation}
here $\tau_{\rm eff}$ is the effective optical depth considering all absorption components (here, we consider FFA optical depth), $T_{\rm CSM}$ is the electron temperature of the CSM, and $\rm{v_{sh}}$ is the shock speed in km\,$\rm s^{-1}$. $\frac{\dot{M}_{-3,\, \rm FFA}}{{\rm v_{CSM,1}}}$ denotes the mass-loss rate in units of $10^{-3}\, \rm M_{\odot}\,yr^{-1}$ for a wind speed of $v_{\rm w} =10\, \rm km\,s^{-1}$. We use $v_{\rm w} =1700\, \rm km\,s^{-1}$ from the P Cygni absorption trough of \ion{He}{1} 5876 \AA\ seen in the optical spectra \citep{Brennan_2024,Dong_2024}.
\par We assume $\mathrm{n(He)/n(H)}\approx 10$ given that the CSM is helium rich \citep{Dong_2024,Brennan_2024}. This ratio could be even higher and the derived mass-loss rate would then be lower. We also assume the CSM temperature to be high at $10^{5}$\,K considering high ionization indicated by the strong presence of helium \citep{Lundqvist_1991}. We note that the $\rm{\dot{M}}$ we estimate will be an effective value, given both that we do not assume an unchanging $K_{2}$ and thus do not fix the mass-loss rate as constant directly, and we assume our line of sight is representative of the total distribution of matter.
 Using the $\mathrm{\tau_{FFA}}$ and the radius of the CSM  $V_{sh}t$ (where we use $V_{sh}=8500$ km/s), we find $\mathrm{\rho_{FFA}}$ at each epoch.

Using the CSM density derived from FFA modeling, we can then constrain $\rm{B}$ using the magnetic field scaling relation \citep{Chevalier_1998}. \begin{equation}\frac{B^2}{8\pi}=\epsilon_{B}\mathrm{\rho_{FFA}}V_{sh}^2
\label{eq:B_from_rho}\end{equation} $V_{sh}$ is again the constant shock speed where $R=V_{sh}t$. 
 Using the estimated $B$ from equation \ref{eq:B_from_rho} and $R$, we can constrain the SSA ${F_{p}}$ and $\rm{\nu_{p}}$  at each epoch using equations \ref{R_eq} and \ref{B_eq}. However, we need $\epsilon_{B}$ and $f_{eB}$. From running MCMC fits using an SSA model attenuated by FFA (equation \ref{eq:SSAFFA}) with $f_{eB}$ and $\epsilon_{B}$ as free parameters, it is clear that there are degeneracies between these values and $\alpha$. However, at the fourth epoch where we have the most datapoints, we find $\epsilon_{B}=0.0016^{+0.0001}_{-0.0001}$ and $f_{eB}=1.2_{-0.2}^{+0.2}$ for the frozen $\alpha$ from the FFA single-epoch fit.  We thus fix $
\epsilon_{B}=0.0016$  and $f_{eB}=1$ given the degeneracy with $\alpha$. This gives good fits as seen in Figure \ref{fig:SSAFFA_Fits} except at the second epoch, where the FFA $\alpha$ and the $f_{eB}$ and $\epsilon_{B}$ that work well for other epochs do not provide a good fit. We attribute this to degeneracies in the fitting. We fix $\alpha=0.75$ at this epoch for the SSA+FFA plotting (this does not affect any derived parameters) given the fit to the optically thin points between 15--30 GHz at all epochs as detailed below. 

We note that we assumed $p \approx 2.5$ in equations \ref{R_eq} and \ref{B_eq}, motivated by the shallow spectral slope derived from an overall fit to the optically thin points between 15--30 GHz (an MCMC fit yields a spectral index of $\alpha = 0.75^{+0.17}_{-0.18}$).   Having found $\rm{\tau_{FFA}}$ and $\rm{F_{p}}/\rm{\nu_{p}}$ through $\rm{\rho_{FFA}}/V_{sh}$ we construct an SSA+FFA model with the $\alpha$ from the single-epoch FFA fits.
\begin{equation}
\begin{split}
F_{\nu,\rm SSA+FFA} = e^{-\mathrm{\tau_{FFA}}} 1.582 F_{p} \left(\frac{\nu}{\nu_{\rm{p}}}\right)^{5/2} \\
\left(1 - \exp\left[-\left(\frac{\nu}{\nu_{\rm{p}}}\right)^{-\alpha - 2.5}\right]\right)
\end{split}
\label{eq:SSAFFA}
\end{equation}
\par 
The 1.582 factor carries over from the SSA formulation \citep{Chevalier_1998}. We did attempt fits where $\tau_{\rm{FFA}}$ is a free parameter along with $\alpha$, $f_{eB}$ and $\epsilon_{B}$. There are clear degeneracies between all these values, so we opted to freeze $\alpha$ and $\tau_{\rm{FFA}}$ from the single-epoch FFA fits and then $f_{eB}/\epsilon_{B}$ from reasonable values obtained from fitting equation \ref{eq:SSAFFA} at the fourth epoch with more datapoints.
We plug in all our values to this equation \ref{eq:SSAFFA} at each epoch. We note that at $>125$ days, there seems to be an uptick in CSM density. Given the assumptions and uncertainties, we suggest these variations are within error bars and cannot be attributed to true variations within the CSM.
As described, we find $\epsilon_{B} = 0.0016$  and $f_{eB}\sim 1$ allows for the best fit to the flux density values, as shown in Figure~\ref{fig:SSAFFA_Fits}. Given the R values we assume and the B values we find in Table \ref{tab:2023fyq_Fitting_results}, we use equation 1 of \citet{Chevalier_1998} to find the density of relativistic electrons $N_{0}$ and thus $U_{E}$. We find $U_{E} \sim 0.8U_{B}$, suggesting that the equipartition implied by our best-fit $f_{eB}=1$ is a reasonable assumption. However, these microparameters are quite difficult to ascertain from the limited data we have and have degeneracies with other parameters i.e., shock radius/$\alpha$ \citep{Frannson_96}. These values differ from the often-assumed $\epsilon_{E} = \epsilon_{B} = 0.1$ \citep{Soderberg_2006}. Recent studies suggest such deviations are plausible due to shock acceleration effects \citep{Park_2015, Gupta_2024}. The inferred value of $\epsilon_{B}$ is also consistent with measurements from other radio-bright interacting SNe such as SN~2023ixf \citep{Nayana_2025}.

\begin{deluxetable*}{ccccccccc}
\tablecolumns{9}
\tabletypesize{\small}
\tablewidth{0pt}
\tablecaption{SN 2023fyq fitting results \label{tab:2023fyq_Fitting_results}}
\tablehead{
    \colhead{Epoch (Days)} & 
    \colhead{$K_1$(FFA)} & 
    \colhead{$K_2$(FFA)} & 
    \colhead{$\rho_{\mathrm{FFA}}$ (g/cm$^3$)} & 
    \colhead{$B$ (G)*} & 
    \colhead{$R$ (cm)*} & 
    \colhead{$\alpha$(FFA)}
}
\startdata
58  & $0.013_{-0.007}^{+0.009}$ & $0.0022_{+0.0003}^{-0.0004}$ & $6.0_{-0.2}^{+0.7} \times 10^{-18}$ & $0.417$ & $4.26 \times 10^{15}$ & $1.49_{-0.70}^{+0.38}$ \\
125 & $0.023_{-0.007}^{+0.010}$ & $0.0025_{-0.0010}^{+0.0010}$ & $1.4_{-0.3}^{+0.4} \times 10^{-18}$ & $0.199$ & $9.18 \times 10^{15}$ & $1.06_{-0.25}^{+0.25}$ \\
139 & $0.017_{-0.004}^{+0.007}$ & $0.0022_{-0.0012}^{+0.0014}$ & $1.0_{-0.2}^{+0.2} \times 10^{-18}$ & $0.174$ & $1.02 \times 10^{16}$ & $0.72_{-0.21}^{+0.24}$ \\
165 & $0.028_{-0.003}^{+0.005}$ & $0.0056_{-0.0010}^{+0.0013}$ & $1.2_{-0.5}^{+0.4} \times 10^{-18}$ & $0.185$ & $1.21 \times 10^{16}$ & $0.78_{-0.06}^{+0.10}$ \\
185 & $0.020_{-0.002}^{+0.003}$ & $0.0030_{-0.0015}^{+0.0019}$ & $6.9_{-5.3}^{+3.6} \times 10^{-19}$ & $0.141$ & $1.36 \times 10^{16}$ & $0.84_{-0.06}^{+0.11}$ \\
\enddata
\tablecomments{Starred values (*) are fixed from the FFA single-epoch fit + assumed shock velocity (using equations \ref{R_eq}/\ref{B_eq} and \ref{eq:B_from_rho}) and thus are not reported with associated error bars. All FFA parameters are from single-epoch fits with equation \ref{eq:FFA} with frozen $\delta=3$/$\beta=1.86$. We fixed $K_{1}<$0.25 at the first epoch due to unphysically large values obtained from fits where it was not constrained.  }
\end{deluxetable*}
\vspace{-0.30 cm}
Our derived SSA+FFA models are plotted in Figure \ref{fig:SSAFFA_Fits} and all relevant parameters are listed in Table \ref{tab:2023fyq_Fitting_results}.  Given we have constrained the frequency $\nu_{p}$ where $\mathrm{\tau_{SSA}}=1$, we find the overall $\mathrm{\tau_{eff}}$ at this frequency using equation \ref{eq:FFA_tau} (adding 1 to the FFA optical depth for the SSA contribution), and estimate the mass-loss rate at each spectral epoch again using equation \ref{eq:Mdot}. Using the 1700 km/s CSM speed, we obtain the mass-loss rates at each epoch with detections seen in Figure \ref{fig:ML}.
The effective mass-loss rate is relatively constant at $\sim 4\times 10^{-3} \ \mathrm{M_{\odot} \ yr^{-1}}$ as expected given the assumed $s=2$ in equation \ref{eq:Mdot}. The variations seen are within error bars. We convert to pre-explosion epoch by taking the ratio of the assumed shock to the CSM speed and plot this as well in Figure \ref{fig:ML}. We discuss the implications of this mass-loss result in Section \ref{sec:Discussion}.  

While we have assumed a standard $r^{-2}$ CSM density profile based on the FFA fitting, SNe Ibn are expected to have an $r^{-3}$ CSM density profile \citep{Maeda_2022}, at least for the first $\sim$ 100 days.
Given the limited data, it is difficult to model in an $s=3$ scenario , especially because at $s=3$ the self-similar solutions derived by \citet{Chevalier_1982} are no longer valid. However, we find the $r^{-3}$ mass-loss rate at each epoch from the relation in \citet{Frannson_96} (equation 2.1 in their work), using our initial mass-loss value as the base mass-loss and adjusting for the evolution of the radius of the shock as needed for non-$\rm{r^{-2}}$ profiles. We also plot this in Figure \ref{fig:SSAFFA_Fits}.

Recent work on stars in binary systems undergoing high rates of mass loss \citep{Wu_2025} has developed an approach that uses a given explosion energy and shock speed as well as an assumed electron power-law index $p$ to obtain estimates of the flux from interaction between the SN ejecta and CSM density profile \citep{Wu_2025}. We display one model from this work in Figure \ref{fig:10GHz} along with the best-fit full (simultaneous multi-epoch using equation \ref{eq:FFA}/\ref{eq:SSA}) SSA/FFA model at 10 GHz. This model uses the M2.467\_P100d $\rho_{\rm CSM} \sim \propto r^{-3}$ density profile (where the density profile is directly extracted from simulations, i.e., no functional form assumed). The model has ejecta kinetic energy of $10^{51}$ erg, ejecta mass 1.25\, M$_{\odot}$, shock velocity of 8970\,km~s$^{-1}$, CSM filling factor 1 (meaning a symmetric CSM), $\epsilon_{e}=10^{-4}$, $\epsilon_{B}=10^{-3}$, and $p=2\alpha+1=2.1$.
\par The shock velocity from this modeling is consistent with our estimate, despite the difference in $p$/small difference in $\epsilon_{e}/\epsilon_{B}$. As shown in Figure \ref{fig:10GHz}, the model does not provide as good a fit to the light curve as SSA/FFA (although data at day $\sim$ 100 would truly have allowed us to distinguish between the models). However, it is noteworthy that our derived density values are relatively consistent (despite our assumption of an $r^{-2}$ profile) with the density profile (the $\mathrm{M2.P467100D}$ model specifically) developed by \citet{Tsuna_2024} for these radio models (e.g., see Fig. \ref{fig:ML}). Furthermore, the \citet{Tsuna_2024} model assumes there is no significant mass-loss beyond 2 $\times 10^{16}$ cm,  which is consistent with our late-time upper limits that suggest a sharp decline in radio flux density.
\subsection{Late-Time Radio/X-ray Non-Detections}
 Our later-time ($\Delta t = 525$\,days) observations at A configuration
 gave stringent $3\sigma$ upper limits on the flux density. We confirmed that these non-detections are not a consequence of the change in array configuration by convolving the A configuration data to match the beam size of the C/D configuration and finding consistent (between the epochs when we had detections and these A configuration observations) flux density for the host and other point sources in the field but not the SN. These non-detections suggest that the CSM density has dropped. 
 To constrain the expected flux density if the mass-loss rate had stayed constant, we extrapolated the FFA parameters derived at $\Delta t = 185$\,days in table \ref{tab:2023fyq_Fitting_results} to $\Delta t = 525$\,days with an $\alpha=0.75$, $\epsilon_{B}=0.0016$ and $f_{eB}=1$. The extrapolated CSM density assuming a similar $\mathrm{K_{2}}$/$\delta$ evolution after day 185 as before day 185 and an $r^{-2}$ decline is $\sim 5\times 10^{-20}\,{\rm g\,cm^{-3}}$.  We plot the extrapolated SSA and SSA+FFA model as well as our non-detections in the final panel in Figure \ref{fig:SSAFFA_Fits}. 
 The extrapolated model exceeds the non-detections by a factor of 2--3 at most frequencies. 
 
 To match the non-detections, as these signify the maximum flux density at that epoch, we find that the density would have to be lower at $\sim 2 \times 10^{-20}\,{\rm g\,cm^{-3}}$ ($\dot{M}=2.5 \times 10^{-3} \rm{M_{\odot}}yr^{-1} $ based on the radius of the CSM at 525 days). We can also find a lower limit on the mass-loss rate in the scenario where the non-detection is solely due to external FFA due to highly dense CSM. We find $\dot{M} \gtrsim 0.8 \ \mathrm{M_{\odot}\,yr^{-1}}$ \citep{Weiler_1986}. 
This seems highly unlikely based on the early values measured, and the needed $\rm{K_{2}}$ would be 3 orders of magnitude higher than the best-fit values at early times (requiring an increase by 3 orders of magnitude in density that would surely have been seen in the X-ray observation at 542 days). We thus do not report any other lower limits as the same logic applies at all other epochs with non-detections.
From the GMRT observation at 256 days, we found $\dot{M}<7 \times 10^{-3} \rm{M_{\odot}\,yr^{-1}}$. This limit is not as constraining as it is roughly the same as prior measured values, and thus we cannot say whether this indicates a change in progenitor mass loss.
 \par We determine an upper limit at early epochs to the mass-loss rate using the same procedure described above, based on the non-detection at 6 days post-explosion. We find a density upper limit of $3 \times 10^{-15} \mathrm{g\,cm^{-3}}$ corresponding to 0.4 $\mathrm{M_{\odot}\,yr^{-1}}$. This is consistent with the mass-loss rate measured with optical models by \citet{Dong_2024}. We add a point from this shock breakout model with CSM interaction developed by \citet{Dong_2024} to Figure \ref{fig:ML} for an early constraint on CSM density. We display all mass-loss and density limits in Figure \ref{fig:ML}. 

With some steepening towards an $r^{-3}$ profile, the late-time non-detection is more consistent with the expected value. It is thus possible that we are actually still observing a CSM with a steeper profile, but in this case, the mass-loss would still have been quite low ($< 10^{-4} \ \rm{M_{\odot}\,yr^{-1}}$at phases $\gtrsim$7 years pre-explosion). 

We also did not detect the supernova at 542 days post-explosion in \textit{Chandra} images, with a limit on the 0.3-10 keV luminosity of $L_{X}<2 \times 10^{38}$ ergs/s. We can convert the limit on luminosity to a limit on mass loss at this epoch. Using the cooling formulas detailed in \citet{Chevalier_17} (equations 14/18), we find from the mass-loss rate inferred from radio modeling ($<$ 0.0025 $\rm{M_{\odot}yr^{-1}}$) that at 542 days, the reverse shock will be adiabatic and there will no longer be significant absorption from the dense shell between the shocks. We thus can use equation 3.10 from \citet{Frannson_96} (which gives the prediction for the 1 keV reverse shock spectral luminosity given some mass-loss rate) to find the mass-loss limit given the limit on the X-ray luminosity. Assuming a 5 keV temperature of the shock, the 8500 km/s shock speed, and converting our X-ray 0.3-10 keV luminosity to a spectral luminosity, we find $\dot{M}<3\times 10^{-3}\rm{M_{\odot}yr^{-1}}$. This is consistent with the limit obtained at radio wavelengths at a similar epoch.

\begin{figure*}
    \centering
    \includegraphics[width=17 cm, height = 8 cm]{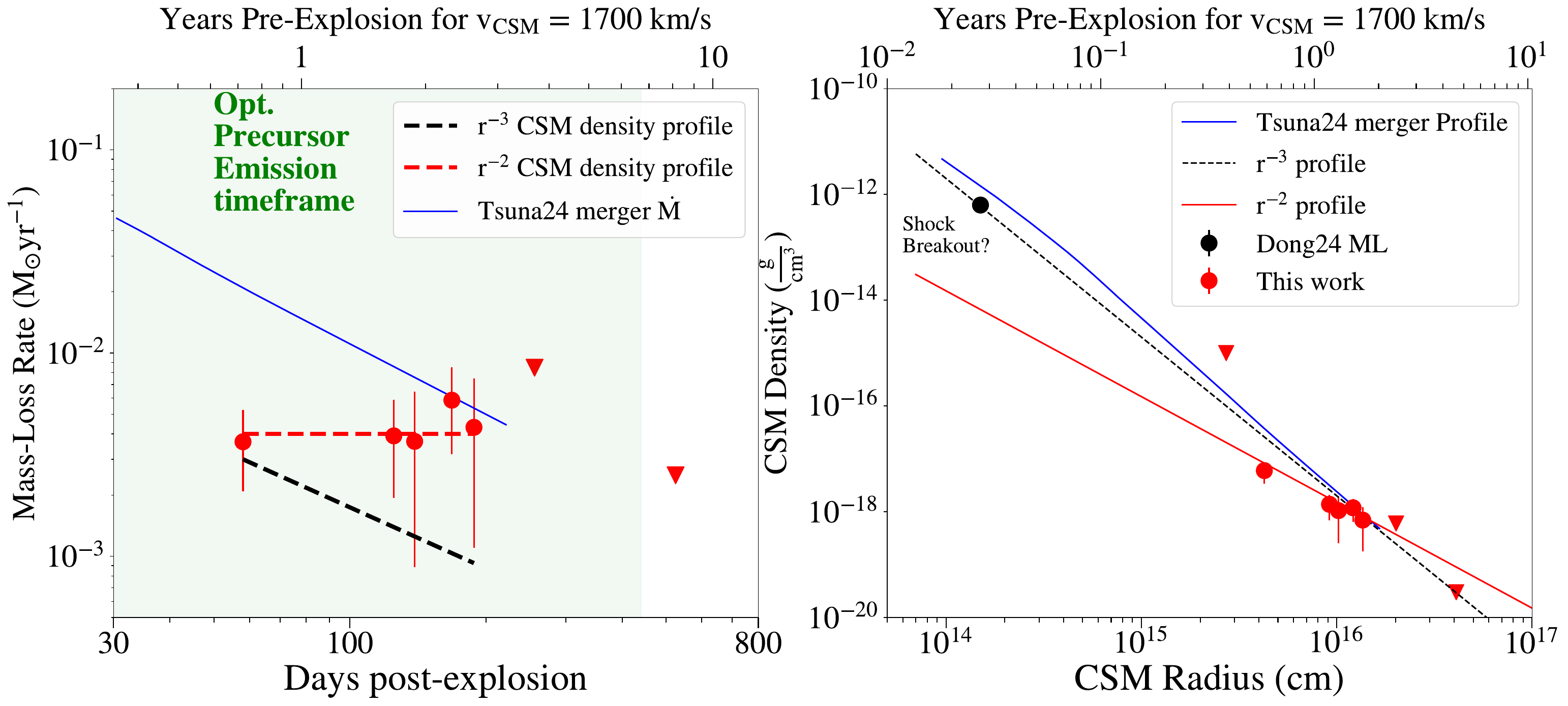}
    \caption{ The mass-loss picture of SN\,2023fyq that we derive based on our radio modeling. $Left:$ Derived effective mass-loss rate as a function of days after explosion, as well as years before explosion, assuming the observed CSM speed of 1700 km/s. We denote the period over which optical precursor emission (at S/N$>$3) was detected \citep{Brennan_2024} to illustrate how the precursor outbursts may have been creating the CSM we observe in the radio. No precursor data could be obtained beyond 7 years before the explosion, since the ZTF/ATLAS surveys had not started then. $Right:$ CSM density associated with the mass-loss rate as a function of the CSM radius. We also show $r^{-3} \ \& \ r^{-2}$ density profiles along with profiles from merger simulations by \citet{Tsuna_2024}. Our data suggests enhanced mass-loss with a sudden drop between 5-8 years before the explosion revealed by our upper limit. We emphasize that some jump in density, potentially due to shock breakout, would be needed to consistently explain our datapoints and the early datapoint from \citep{Dong_2024} in the $r^{-2}$ case. Our last $3\sigma$ upper limits are consistent with the \citet{Tsuna_2024} density profile extrapolation, given that this profile assumes that the mass loss starts directly at 4 years before the explosion (for our assumed CSM speed).}
    \label{fig:ML}
\end{figure*}
\section{Discussion}\label{sec:Discussion}
The detection of radio emission from SN 2023fyq allows for the first constraints on mass-loss up to 8 years pre-explosion in a SN\, Ibn.
There are two clear results from our modeling:
\begin{enumerate}
    \item The best-fit model to the radio emission suggests mass loss at a rate $\sim$0.004\,M$_{\odot}\,\mathrm{yr}^{-1}$ from 0.7--3 years before the explosion for a CSM velocity of 1700 km/s as measured in optical spectra \citep{Dong_2024}. This mass-loss rate translates to constraints on CSM density from (0.4--1.4)$\times 10^{16}$ cm.
    \item The non-detections at $\Delta t = 525$\, days suggest that there was a drop in CSM density at $R \sim 2\times 10^{16}$ cm (suggesting a drop by a factor $>$ 2 in progenitor mass-loss rate). This is confirmed by an X-ray non-detection at $\Delta t = 542$\,days, which gives a similar limit on the mass-loss rate and implies a shell-like CSM.
\end{enumerate}
\par From an analysis of the optical spectroscopy and lightcurves, \cite{Dong_2024} and \cite{Brennan_2024} determined that SN\,2023fyq was surrounded by a highly dense torus of CSM created by intensive mass loss (at a rate of $\sim$ 1 $\mathrm{M_{\odot}\,yr^{-1}}$) in the last 40 days before explosion that caused the sharply rising precursor emission. Assuming our CSM speed of 1700 km/s, this would only create CSM out to $R \approx 3 \times 10^{15}$ cm, which is a smaller radius than the CSM radius at any epoch at which we obtained detections. The authors also suggest that there was further mass loss at earlier times before the explosion caused by less intensive precursor outbursts that may have created a shell of less dense CSM (with no specifics on the density) at larger radii ($R \gtrsim 3 \times 10^{15}$\,cm). 
This shell could correspond to the material traced by our radio observations, whereas the mass loss immediately preceding the explosion may have caused strong radio absorption consistent with our early non-detections up to $\Delta t = 58$\,days ($R \lesssim 4\times10^{15}$\,cm).
\par On the nature of the progenitor, \citet{Dong_2024} concluded that SN\,2023fyq was a helium star in a binary system with a compact object which either merged or experienced runaway core collapse very close to explosion. This merger-precursor scenario has been considered and modeled in detail by \citet{Tsuna_2024}, and the mass-loss rate we measure in these initial phases up to 1 year pre-explosion is generally consistent with the mass-loss values they find. However, it does not match the increase in mass-loss rate (as we assume a standard $r^{-2}$ profile given that our data do not fully constrain the density profile) as seen in Figure \ref{fig:ML}. This short mass-loss timescale ($<$ 10 years before the explosion) is common in type Ibn supernovae \citep{Smith_2014}, and the merger scenario does provide a clear explanation in this case for why the mass-loss would only occur at an elevated rate for such a limited time frame. 

In more detail, the sharp drop in CSM density at 2$\times 10^{16}$ cm is only feasible in a scenario where there were limited mass ejections or some late-stage mass loss explicitly triggered by changes in the progenitor system. The merger model can account for the uptick in mass loss, with the beginning of mass-loss caused by the onset of non-conservative mass transfer preceding merger \citep{Tsuna_25,Tsuna_2024}. The mass-loss rate measured from 0.7-3 years at 0.004 $\rm{M_{\odot}yr^{-1}}$ is, in fact, relatively low (by an order of magnitude) compared to the huge outbursts from mass ejections seen in SN 2006jc or 2022ablq, which may suggest that the merger model for SN 2023fyq is more likely than mass ejections. The elevated mass-loss just before the explosion, suggested by the rise in precursor emission seen at optical wavelengths in the last 40 days, also means that there would have to have been multiple outbursts at varying mass-loss rates in the ejection scenario. Additionally, the consistent optical emission over $>$ 5 years suggests continuous energy injection and perhaps disfavors the ejection scenario \citep{Dong_2024}. Given the extremely short timescale and the high mass-loss rate, a stellar wind cannot explain our measurements. However, our data cannot explicitly distinguish between distinct ejections leading to the optical/radio discrepancy at different times or a continuous outflow driven by the merger inspiral, given the cadence of our data/uncertain constraints on density profile. We also note that our data does not fully distinguish between our preferred CSM shell vs. torus/clumps; however, clumps would be expected to give rise to elevated X-ray emission/a potentially shallower density profile that we do not see. 

The timescale of mass loss in SN\,2023fyq differs from other SNe\,Ibn with confirmed multi-wavelength emission that allowed for a constraint on the mass-loss evolution across time. In SN\,2006jc, detailed modeling of the X-ray light curve and the optical data revealed a massive outburst two years before explosion at $\dot{M} \sim 10^{-2} \mathrm{M_{\odot}yr^{-1}}$ that was preceded by extremely quiescent mass-loss at $>$ 5 orders of magnitude lower \citep{Tominaga_2008}. SN\,2022ablq revealed even more intensive mass loss at $\dot{M} \sim 10^{-1}~\mathrm{M_{\odot}yr^{-1}}$ for the 2--3 years pre-explosion \citep{Pellegrino_2024}.
\par While the mass-loss rate has ramped up pre-explosion in SN~2023fyq as shown by the optical analysis \citep{Dong_2024,Brennan_2024}, the radio results suggest the increase was only by $\approx$ 2 orders of magnitude. It thus seems that the mass loss acted on a different timescale and at a different magnitude in SN\, 2023fyq, emphasizing that SNe Ibn can originate from a wide range of progenitor channels. While initially it was thought that most SNe Ibn came from Wolf-Rayet stars \citep{Pastorello_2016}, certain objects showed potential evidence for more exotic progenitors such as white dwarf binary systems \citep{Sanders_2013}. Other work has suggested that certain pair instability supernovae could appear as SNe\,Ibn with hydrogen-free ejecta \citep{Woosley_2017,Renzo_2020} due to the intense mass ejections. 
\par
 SN\,2023fyq is the first type Ibn object to be detected in the radio, but it is not the first object for which observations were attempted. SNe\, Ibn 2019aajs and SN \,2019qav were observed but not detected with the VLA and NOEMA (IRAM Northern Extended Millimeter Array) from 15-400 days post-explosion at 10/90 GHz \citep{Ho_2023}. SN\,2020bqj was observed but not detected with AMI-LA at 15.5 GHz at 6 days post-explosion \citep{Kool_2021}. SN\,2015G \citep{Shivvers_2017} and the aforementioned SN\,2006jc \citep{Soderberg_2006} were both observed at VLA frequencies from 4--22 GHz at$\sim$ 1 week for SN\, 2006jc and out to 120 days for SN\, 2015G. The dates of discovery/explosion are taken from optical estimates for all of these SNe.
 
 For SN\,2015G, the non-detection was analyzed to obtain a limit on the mass-loss rate at $< 10^{-4} \,\mathrm{M_{\odot}\,yr^{-1}}$, while for SNe\,2006jc, 2019aajs, 2019qav and 2020bqj the results were less constraining, given that the non-detections were either early at 7-40 days post-explosion \citep{Pastorello_2008,Kool_2021,Ho_2023} or were at a high luminosity limit given the large distance to the SN. SN\,2015G had a 10 GHz non-detection at $3\times 10^{25}\rm{erg\,s^{-1}Hz^{-1}}$ at 120 days, which is a factor of 5 below the luminosity of the 10 GHz detection of SN\, 2023fyq at 125 days. We display the radio light curve of SN~2023fyq and the upper limits for the other SNe~Ibn in Figure \ref{fig:Radio_context}. 
The fact that the SN~2015G radio emission must have been at least one order of magnitude less luminous implies that 2015G may come from a progenitor with a different mass-loss history.
\par When comparing with other interacting SNe, SN\,2023fyq is less luminous and peaks slightly earlier than most radio-detected SNe~IIn as shown in Figure \ref{fig:23fyq_Lpnup}. Another difference from SNe~IIn is in the non-detections at later times and/or potentially steeper density profile. While SN 2023fyq may have experienced mass loss for longer than other SNe Ibn, it appears that it definitely did not lose mass at enhanced rates on the decade-centuries timescales that SNe IIn do (see i.e. \citet{Smith_2016}). SN\,2023fyq provides compelling evidence for continuing to observe a larger sample of SNe Ibn at radio wavelengths to try to understand the subclass.
 \begin{figure}[H]
    \centering
    \includegraphics[width=7 cm, height= 9 cm]{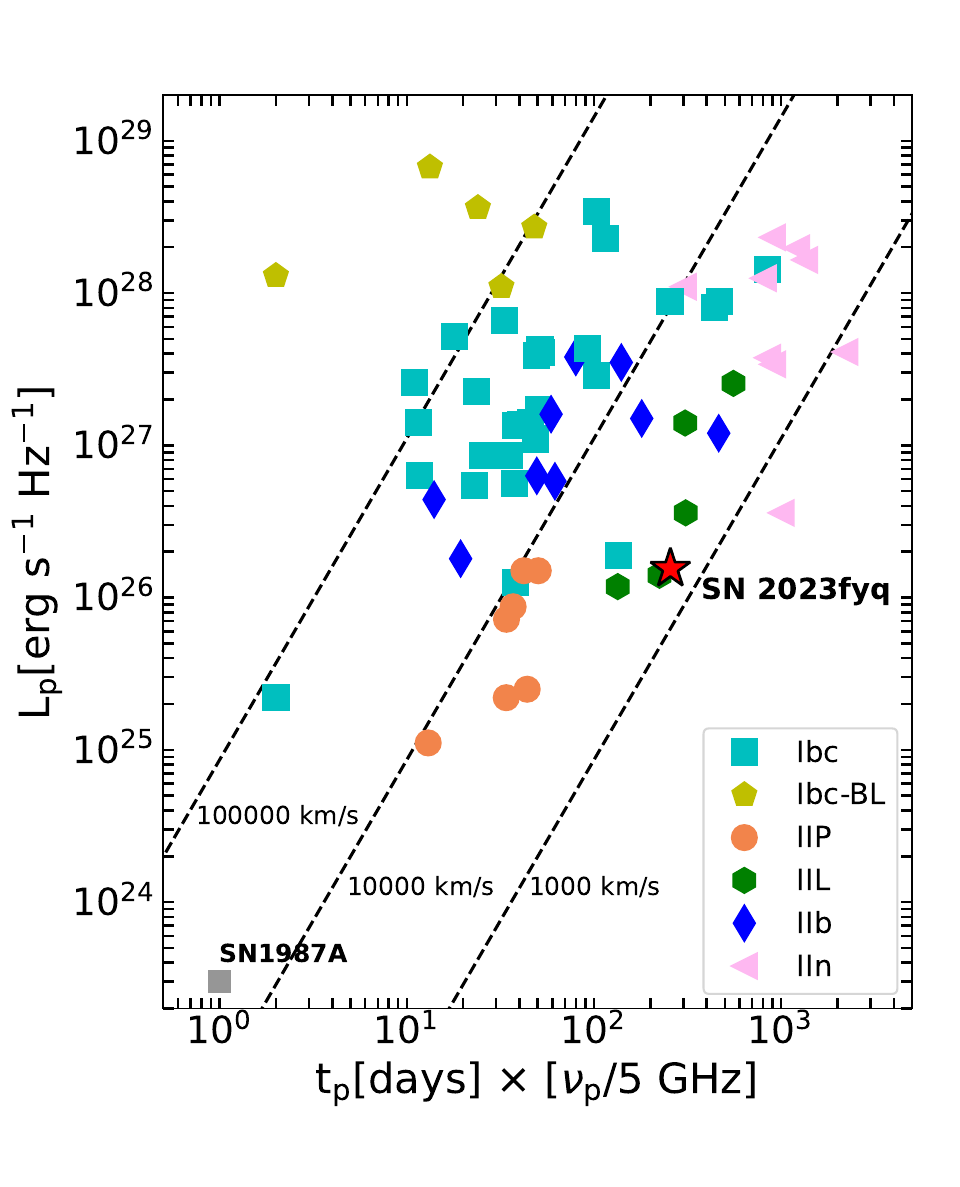}
    \caption{The peak spectral luminosity vs. peak time for SN 2023fyq (taken at 10 GHz) in context with a variety of other radio-detected SNe (\citet{Nayana_2025} and references therein). The speeds noted on diagonal lines are shock velocities for pure SSA. 2023fyq is distinct from most other stripped-envelope SNe, likely due to denser CSM, but does not reach the peak spectral luminosity of many SNe ~IIn.}
    \label{fig:23fyq_Lpnup}
\end{figure}

\section{Conclusion}\label{sec:Conclusions}

We observed SN\,2023fyq from 1--35 GHz with the VLA and GMRT over 1.5 years post-explosion.  We also analyzed data over the same time period at X-ray wavelengths from \textit{Swift-XRT} and \textit{Chandra} (finding non-detections). Our detections at radio wavelengths constitute the first ever radio detection and long-term radio follow-up of a type Ibn supernova. We obtain constraints on the mass-loss rate from the best-fitting free-free absorbed + intrinsic synchrotron self-absorbed models. The magnitude of the mass-loss rate and CSM density/upper limits measured suggest a drop in CSM density at $\sim$ $2\times 10^{16}$ cm that can be explained by enhanced ($\sim 4\times10^{-3}~\mathrm{M_{\odot}\,yr^{-1}}$) mass loss that only occurred for roughly 5 years before the explosion and must have ramped up by $\sim$ 2 orders of magnitude in the final months pre-explosion based on the optical evolution \citep{Brennan_2024,Dong_2024}.
Our derived mass-loss history is roughly consistent with early mass-loss predictions from the binary merger scenario explored in depth by \cite{Tsuna_2024}. We emphasize that further monitoring of SNe~Ibn at radio wavelengths within 1-2 years post-explosion is needed to constrain their progenitor systems and understand objects like SN\,2023fyq in the context of the larger SN~Ibn population. 
\begin{figure}[H]
    \centering
    \includegraphics[width=7.5 cm, height=6 cm]{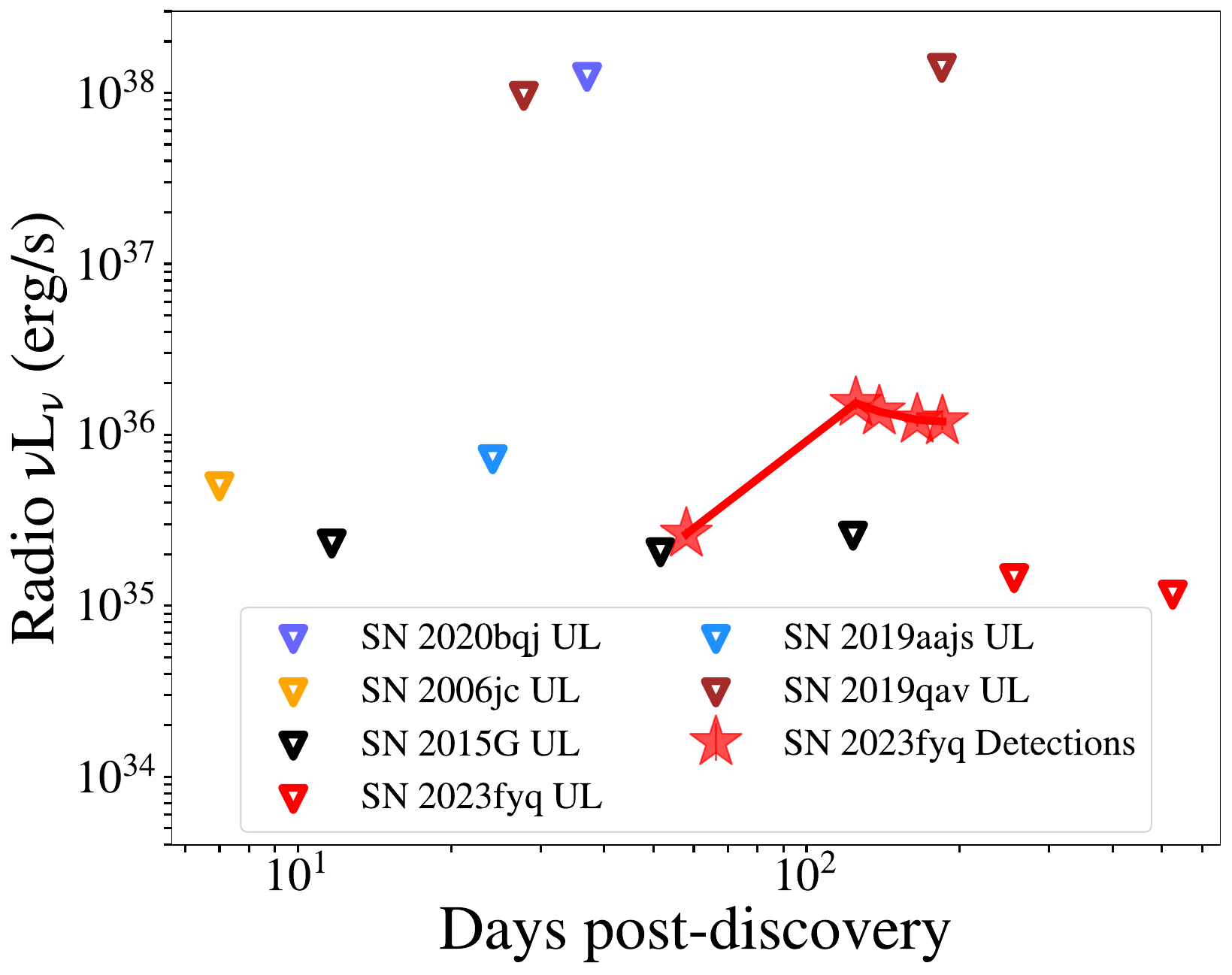}
    \caption{A view of all previous radio-observed type Ibn SNe and the upper limits on luminosities, including 10 GHz detections and non-detections for SN\, 2023fyq. Non-detections are shown with downward-facing triangles, and are all 3$\sigma$ limits. SNe 2006jc, 2019qav, 2019aajs and 2020bgq were observed quite early and thus their radio evolution could still be consistent with SN 2023fyq, but SN~2015G clearly shows upper limits that are much deeper than the observed radio emission from  SN~2023fyq. Future deep observations between 50-500 days post-explosion will be key to constraining the subclass. }
    \label{fig:Radio_context}
\end{figure}
\facilities{GMRT, VLA, Chandra, Swift (XRT)}
\\
\textit{Software:} \texttt{Astropy} \citep{Astropy_2013,Astropy_2022,Astropy_2018}, CASA \citep{Casa_desc}, \texttt{xspec}\citep{xspec}, CIAO\citep{Ciao_2006}
\section{Acknowledgments}
We thank the referee for insightful comments that have improved this manuscript greatly.
RBW is supported by the National Science Foundation Graduate Research Fellowship Program under Grant number 2234693 and acknowledges support from the Virginia Space Grant Consortium.
M.M. and the METAL group acknowledge support in part from ADAP program grant No. 80NSSC22K0486, from the NSF grant AST-2206657 and from the National Science Foundation under Cooperative Agreement 2421782 and the Simons Foundation grant MPS-AI-00010515 awarded to the NSF-Simons AI Institute for Cosmic Origins — CosmicAI, https://www.cosmicai.org/.
The Very Large Array is operated by the National Radio Astronomy Observatory, a facility of the U.S. National Science Foundation (NSF) operated under cooperative agreement by Associated Universities, Inc. 
The GMRT is run by the National Centre for Radio Astrophysics of the Tata Institute of Fundamental Research. 
This work has used data from the Chandra Data Archive and software provided by the Chandra
X-ray Center (CXC) in the application package CIAO.
D.T. is grateful for support from the Sherman Fairchild Postdoctoral Fellowship at Caltech. 
W.J.-G. is supported by NASA through Hubble Fellowship grant HSTHF2-51525.001-A awarded by the Space Telescope Science Institute, which is operated for NASA by the Association of Universities for Research in Astronomy, Inc., under contract NAS5-26555.
M.R.D. acknowledges support from NSERC through grant RGPIN-2019-06186, the Canada Research Chairs Program, and the Dunlap Institute at
the University of Toronto. 
R.M. acknowledges support by the National Science
Foundation under award No. AST-2224255.
C.D.K. gratefully acknowledges support from the NSF through AST-2432037, the HST Guest Observer Program through HST-SNAP-17070 and HST-GO-17706, and from JWST Archival Research through JWST-AR-6241 and JWST-AR-5441.
\newline

\begin{deluxetable*}{ccccccc}
\tablecolumns{7}
\tabletypesize{\small}
\tablewidth{\textwidth}
\setlength{\tabcolsep}{3pt}
\tablecaption{The radio dataset obtained of SN\,2023fyq\label{tab:2023fyq}.}
\tablehead{
    \colhead{Telescope} & \colhead{Date} & \colhead{Epoch (Days since explosion)} & \colhead{Freq (GHz)} & \colhead{Flux Density (mJy)} & \colhead{Image RMS ($\mu$Jy)}
}
\startdata
VLA & 07/22/2023 & 6 & 10.0 & $<$0.018 & 6 \\ 
VLA & 08/08/2023 & 23 & 10.0 & $<$0.015 & 5 \\ 
GMRT & 08/13/2023 & 28 & 1.25 & $<$0.45 & 150 \\ 
VLA & 09/12/2023 & 58& 11.0 & 0.092 $\pm$ 0.011 & 11 \\ 
VLA & 09/12/2023 & 58 & 9.0 & 0.0416 $\pm$ 0.004 & 7 \\ 
VLA & 09/12/2023 & 58 & 10.0 & 0.068 $\pm$ 0.005 & 6 \\ 
VLA & 11/18/2023 & 125 & 3.0 & $<$1.5 & 100 \\ 
VLA & 11/18/2023 & 125 & 6.0 & 0.396 $\pm$ 0.032 & 30 \\ 
VLA & 11/18/2023 & 125 & 8.1 & 0.415 $\pm$ 0.036 & 60 \\ 
VLA & 11/18/2023 & 125 & 10.0 & 0.387 $\pm$ 0.035 & 25 \\ 
VLA & 11/18/2023 & 125 & 15.0 & 0.315 $\pm$ 0.005 & 15 \\ 
VLA & 11/18/2023 & 125 & 20.0 & 0.244 $\pm$ 0.012 & 22 \\ 
VLA & 11/18/2023 & 125 & 24.0 & 0.206 $\pm$ 0.015 & 25 \\ 
VLA & 12/02/2023 & 139 & 3.0 & $<$1.1 & 60 \\ 
VLA & 12/02/2023 & 139 & 6.0 & 0.353 $\pm$ 0.018 & 23 \\ 
VLA & 12/02/2023 & 139 & 8.1 & 0.321 $\pm$ 0.024 & 60 \\ 
VLA & 12/02/2023 & 139 & 10.0 & 0.339 $\pm$ 0.007 & 25 \\ 
VLA & 12/02/2023 & 139 & 15.0 & 0.29 $\pm$ 0.029 & 20.0 \\ 
VLA & 12/02/2023 & 139 & 20.0 & 0.249 $\pm$ 0.015 & 30 \\ 
VLA & 12/02/2023 & 139 & 24.0 & 0.204 $\pm$ 0.012 & 25 \\
VLA & 12/28/2023 & 165 & 3.0 & $<$1.2 & 100 \\ 
VLA & 12/28/2023 & 165 & 8.1 & 0.326 $\pm$ 0.0097 & 20.0 \\
VLA & 12/28/2023 & 165 & 6.0 & 0.316 $\pm$ 0.016 & 50 \\ 
VLA & 12/28/2023 & 165 & 10.0 & 0.315 $\pm$ 0.029 & 30 \\ 
VLA & 12/28/2023 & 165 & 15.0 & 0.297 $\pm$ 0.021 & 25 \\ 
VLA & 12/28/2023 & 165 & 20 & 0.246 $\pm$ 0.014 & 35 \\ 
VLA & 12/28/2023 & 165 & 24.0 & 0.22 $\pm$ 0.014 & 35 \\ 
VLA & 12/28/2023 & 165 & 31.0 & 0.208 $\pm$ 0.02 & 30 \\ 
VLA & 12/28/2023 & 165 & 35.0 & 0.189 $\pm$ 0.018 & 30 \\ 
VLA & 01/17/2024 & 185 & 3.0 & $<$0.471 & 30 \\ 
VLA & 01/17/2024 & 185 & 6.0 & 0.309 $\pm$ 0.017 & 55 \\ 
VLA & 01/17/2024 & 185 & 8.1 & 0.276 $\pm$ 0.017 & 60 \\ 
VLA & 01/17/2024 & 185 & 10 & 0.2 $\pm$ 0.034 & 25 \\ 
VLA & 01/18/2024 & 186 & 15.0 & 0.18 $\pm$ 0.015 & 50 \\ 
VLA & 01/18/2024 & 186 & 20.0 & 0.142 $\pm$ 0.012 & 18 \\ 
VLA & 01/18/2024 & 186 & 24.0 & 0.135 $\pm$ 0.005 & 15 \\ 
VLA & 01/18/2024 & 186 & 31.0 & 0.116 $\pm$ 0.008 & 20.0 \\ 
VLA & 01/18/2024 & 186 & 35.0 & 0.108 $\pm$ 0.013 & 25 \\
GMRT & 03/28/2024 & 256 & 1.25 & $<$0.3 & 100 \\  
VLA & 12/23/2024 & 525 & 3.0 & $<$0.042 & 14 \\ 
VLA & 12/23/2024 & 525 & 6.0 & $<$0.03 & 10.0 \\ 
VLA & 12/23/2024 & 525 & 10.0 & $<$0.03 & 10.0 \\ 
VLA & 12/23/2024 & 525 & 15.0 & $<$0.039 & 9.5 \\
\enddata
\tablecomments{The VLA data were obtained under program 23A-157 (PI W.~Jacobson-Galan) and 24B-500 (PI R.~Baer-Way). The GMRT data were taken under GMRT proposal $45\_091$ in cycle 45 (PI P.~Chandra). Upper limits are 3$\sigma$. The reported errors are errors directly from \texttt{imfit}.}
\end{deluxetable*}
\vspace{5mm}




\let\cleardoublepage\clearpage

\

\bibliography{ms}{}
\bibliographystyle{aasjournal}



\end{document}